# Astrobiological significance of minerals on Mars surface environment: UV-shielding properties of Fe (jarosite) vs. Ca (gypsum) sulphates

Gabriel Amaral[1], Jesus Martinez-Frias[2], Luis Vázquez[2,3]

[1] Departamento de Química Física I, Facultad de Ciencias Químicas, Universidad Complutense, 28040 Madrid, Spain; Tel: 34-91-394-4305/4321, Fax: 34-91-394-4135, e-mail: gamaral@quim.ucm.es

[2] Centro de Astrobiología (CSIC-INTA), 28850 Torrejón de Ardoz, Madrid, Spain, Tel: 34-91-520-6418, Fax: 34-91-520-1621, e-mail: martinezfrias@mncn.csic.es

[3] Departamento de Matemática Aplicada, Facultad de Informática, Universidad Complutense de Madrid, 28040 Madrid, Spain, Tel: +34-91-3947612, Fax: +34-91-3947510, e-mail: lvazquez@fdi.ucm.es

**Corresponding author:** Jesús Martinez-Frias (martinezfrias@mncn.csic.es)

**Abstract**

The recent discovery of liquid water-related sulphates on Mars is of great astrobiological interest. UV radiation experiments, using natural Ca and Fe sulphates (gypsum, jarosite), coming from two selected areas of SE Spain (Jaroso Hydrothermal System and the Sorbas evaporitic basin), were performed using a Xe Lamp with an integrated output from 220 nm to 500 nm of 1.2 Wm$^{-2}$. The results obtained demonstrate a large difference in the UV protection capabilities of both minerals and also confirm that the mineralogical composition of the Martian regolith is a crucial shielding factor. Whereas gypsum showed a much higher transmission percentage, jarosite samples, with a thickness of only 500 µm, prevented transmission. This result is extremely important for the search for life on Mars as: a) jarosite typically occurs on Earth as alteration crusts and patinas, and b) a very thin crust of jarosite on the surface of Mars would be sufficient to shield microorganisms from UV radiation.

## 1. Introduction

Over the last half century, Mars has been explored with telescopes, spacecrafts and robotic rovers. All the information yielded from these different sources, along with the results obtained by the study of SNC meteorites and terrestrial analogs, is starting to reveal the geological diversity of the planet and provides data for theorizing about how the different Martian environments evolved. However, despite the huge amount of geomorphological, geodynamic and geophysical data obtained, much is still unknown about Mars mineralogy and paragenetic assemblages, which is fundamental to an understanding of its whole geological history. Minerals are not only indicators of the physical-chemical settings of the different environments and their later changes, but also they could (and can) play a crucial astrobiological role related with the possibility of existence of extinct or extant Martian life. This paper aims: 1) to emphasize the significance of Mars minerals as environmental geomarkers for the search for extraterrestrial life and 2) to present, for the first time, a comparative study of the UV-shielding properties of two sulphate minerals (jarosite ($KFe_3(SO_4)_2(OH)_6$) and gypsum ($CaSO_4.2H_2O$)). Both sulphates were recently found to exist on the UV-radiation, hostile environmental conditions of the Mars surface and their formation was related with liquid water.

## 2. Mineralogy and UV-radiation on the surface of Mars

Information from scientific literature about past Mars missions, together with recent reviews and new findings (see for instance Souza et al. 2004,

Mcsween 2004, Squyres and Knoll, 2005, Clark et al. 2005, Poulet et al. 2005) indicate that the Mars surface environment is dominated by minerals characteristic of mafic igneous rocks (although controversy still remains about the existence of andesite versus chemically weathered basalts). Major surface geological units of the ancient crust consist of pyroxenes and plagioclase, with varying proportions of olivine and alteration minerals. Quartzofeldspathic materials also have been identified (Bandfield et al. 2004). The Martian regolith is made up of an apparently homogenized dust having basaltic composition, with admixed local rock components, oxides, water-bearing phyllosilicates and salts (mainly sulphates). As well, Martian (SNC) meteorites display small amounts of secondary minerals (clays, carbonates, halides, sulphates) probably formed by reaction with subsurface fluids.

Unlike Earth, there is a significant amount of UV flux on Mars, mainly due to the influence of the shorter wavelengths UVC (100-280 nm) and UVB (280-315 nm). Various works on the biological effects of UV radiation (Cockell, 1998, Cockell et al. 2000) have established that even the present-day instantaneous Martian UV flux would not in itself prevent life. Nevertheless, it is a fact that this UV flux contributes, coupled with the lack of liquid water and extreme low temperatures, to the biologically inhospitable nature of the Martian surface. From the astrobiological point of view, these factors render a practical consequence for the exploration and detection of life on Mars: any living organism, as we know it, should have preferentially developed in a particular sub-surface microenvironment able to protect it from the harsh conditions on the surface. Terrestrial endolithic communities that live in the subsurface layers of rock that provide appropriate microenvironments against extreme external conditions have been proposed (Friedmann, 1982, McKay, 1993, Wynn-Williams and Edwards, 2000, Villar et al. 2005) as possible analogs to life on Mars. Very probably no place on Earth is truly like Mars. Nevertheless, it is possible to define potential sites on our planet (Mars analogs) where environmental conditions (geology, tectonics, *mineralogy*, etc) approximate, in some specific way, those possibly encountered on Mars at present or earlier in that planet's history. Extant Martian life would require strong UV shielding, which, in accordance with the results here presented, could be perfectly accomplished by certain minerals already discovered on Mars.

*2.1 Minerals as environmental geomarkers*

The search for life on Mars has been intimately linked to the identification of unequivocal ancient or modern geomarkers of water on and inside the planet. Recently, some water-related sulphates (e.g. jarosite, gypsum) were discovered (Squyres et al. 2004, Klingelhöfer et al. 2004, Langevin et al. 2005) on Mars surface. In particular, jarosite (Figure 1a) has proven to have a great astrobiological importance, not only for its relation with liquid water, but also because it can act as a sink and source of Fe ions for Fe-related chemolithoautotrophic microorganisms, such as those encountered in numerous extremophilic ecosystems (e.g. Tinto river (Lopez-Arcilla et al. 2001, Gonzalez-Toril et al. 2003, Amaral Zettler et al. 2003, Fernandez-Remolar et al. 2004, 2005).

Jarosite is a mineral of the alunite-jarosite family. In accordance with Scott (2000), the alunite–jarosite minerals are defined as having the general formula $AB_3(XO_4)_2(OH)_6$, where A is a large ion in 12-fold coordination (e.g., K, Na, Ca, Pb, REE), B is usually Fe or Al, and the $XO_4$ anions are usually $SO_4$, $PO_4$ or $AsO_4$. Gypsum, $CaSO_4.2H_2O$, is essentially a layered structure bound by hydrogen bonds. Zig-zag chains of $CaO_8$ polyhedra, running parallel to c, are bound together by similar chains of isolated $(SO_4)^{2-}$ tetrahedra, forming a double sheet perpendicular to (010). Each $Ca^{2+}$ ion is surrounded by six oxygen atoms belonging to the sulphate groups and two oxygen atoms belonging to the $H_2O$ molecules. These $H_2O$ molecules form a layer binding the polyhedral sheets together with weak hydrogen bonds. The $H_2O$ molecules are significantly distorted and are oriented such that the hydrogen bond H2····O1 acts almost entirely along b (Schofield et al. 1996).

The samples of jarosite and gypsum used in this work come from two selected areas of the SE Mediterranean region of Spain: Jaroso and Sorbas, which have been proposed as a relevant geodynamic and mineralogical model (Martinez-Frias et al. 2001a, 2001b, Martinez-Frias, 2004, Rull et al. 2005) to follow for the astrobiological exploration of Mars. The Jaroso Hydrothermal System is a volcanism-related multistage hydrothermal episode of Upper Miocene age, which includes oxy-hydroxides (e.g. hematite), base- and precious-metal sulfides and different types of sulphosalts. Hydrothermal fluids and sulfuric acid weathering of the ores have generated huge amounts of oxide and sulfate minerals of which jarosite is the most abundant (Martinez-Frias et al. 1992, Martinez-Frias, 1998).

It is important to note that jarosite was first discovered on Earth at this area, in 1852, in the "Jaroso Ravine", which is the world type locality of jarosite (Amar de la Torre, 1852, Martinez-Frias, 1999). The Sorbas basin contains one of the most complete sedimentary successions of the Mediterranean (gypsum karst) reflecting the increasing salinity during the Messinian salinity crisis (desiccation of the Mediterranean Sea) (Martin and Braga, 1994, Riding et al. 1998, Krijgsman et al. 1999) and showing a complex paleogeographical evolution, being a signature of its progressive restriction and isolation.

*2.2 Materials and methods*

Jarosite samples were pulverized and shaped by pressure into flat round pellets of 1 cm diameter and 0.5 mm thickness (Figure 1b). Gypsum samples were scratched from a larger specimen. The samples were flattened to different thickness (between 0.1 mm to 1.6 mm) before exposed to UV light. All samples were positioned onto a sample holder which allowed the UV lamp light to go through the samples and be detected by a PMT through a monochromator.

The UV light source is a Xe Lamp (SpectralProducts) with an integrated output from 220 nm to 500 nm of 1.2 $Wm^{-2}$ (measured at a distance of 10 cm from the lamp exit and over a disc of 5 mm diameter). A monochromator (SpectralProducts, 1/8 m CM 110) is placed at the output of the lamp, but in our experiments it is set up so that it lets all wavelenghts out without deformation of the Xe lamp output. 5 cm downstream the light beam irradiates the sample, the light which is transmitted through it is then collected by an optical fiber head (placed 10 cm away from the Xe lamp), which in turns directs the light through through a spectrometer (Bentham, CM 150 double monochromator) which records the transmitted light spectrum. The light is detected by a peltier-cooled bi-alkali photomultiplier tube (PMT Bentham, DH-10-Te). Typically, each sample spectrum is measured from 220 nm to 500 nm every 1nm with a rate of approximately 1 nm/s.

Usually each sample spectrum is the average of at least 5 spectra. The mineralogical and chemical composition of both sulphates was previously verified by XRD (X-ray diffraction), scanning electronic microscopy (SEM-EDX) and ICP-MS (Inductively Coupled Plasma-Mass Spectrometry).

## 3. Results and conclusion

It is well known that iron and iron-bearing compounds can provide an UV screen for life. For instance, Martian regolithic dust, along with palagonite-type compositions, have specifically been suggested as a possible safe haven for life (Sagan and Pollack, 1974, Olsen and Pierson, 1986, Pierson et al. 1993, Kumar et al. 1996, Allen et al. 1998, Phoenix et al. 2001, Gomez et al. 2003, among others). Our experimental results demonstrate a large difference in the UV shielding capabilities of both sulphates (Figure 2). Whereas gypsum showed a much higher transmission percentage (with an error in the absorption coefficients of roughly 20%), jarosite samples, with a thickness of only 500 µm, prevented transmission (Figure 2). This has great astrobiological relevance as: a) jarosite typically occurs on Earth as alteration crusts and patinas (see figure 1a), and b) a very thin crust of jarosite on the surface of Mars would be sufficient to shield microorganisms from UV radiation.

**Acknowledgements**

This work was supported by the Spanish Centro de Astrobiologia (CSIC/INTA), associated to the NASA Astrobiology Institute. Thanks to the Rover Environmental Monitoring Station (REMS) project. Maite Fernandez Sampedro, Maria Paz Martín Redondo and Dr. Virginia Souza-Egipsy are acknowledged for their assistance with the analyses. Special thanks to Dr. David Hochberg for the revision of the English version.

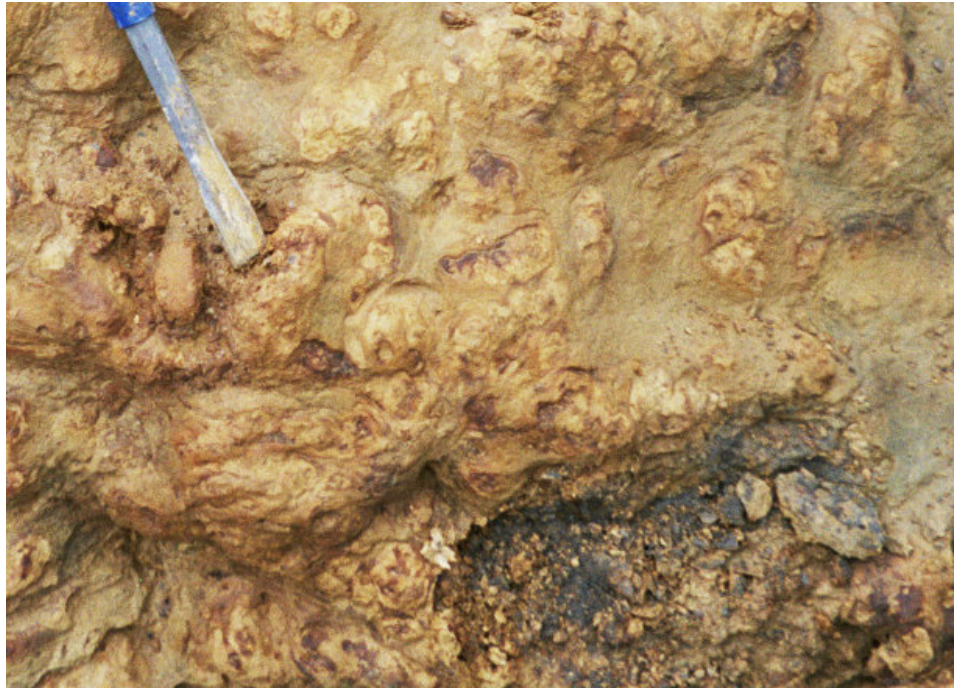

Figure1a: Typical alteration crust rich in jarosite at El Jaroso ravine, Cuevas del Almanzora Natural Area, Almería province, SE Spain.

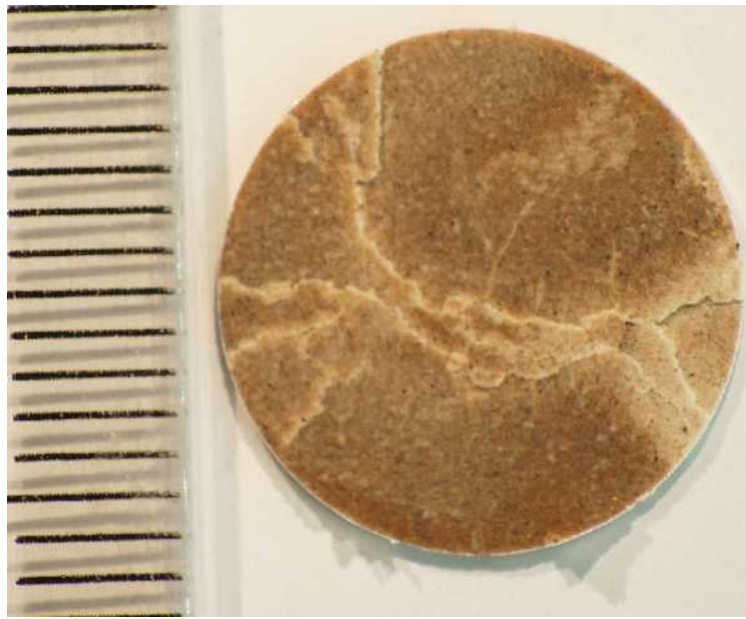

Fig 1b: Jarosite pellet. Each division on the left scale represents 1 mm.

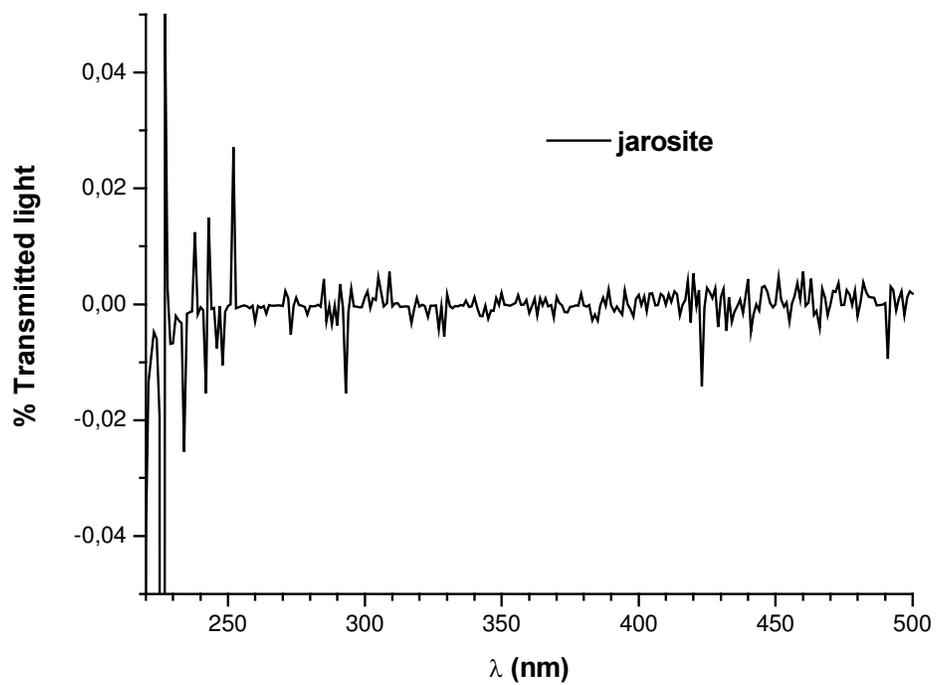

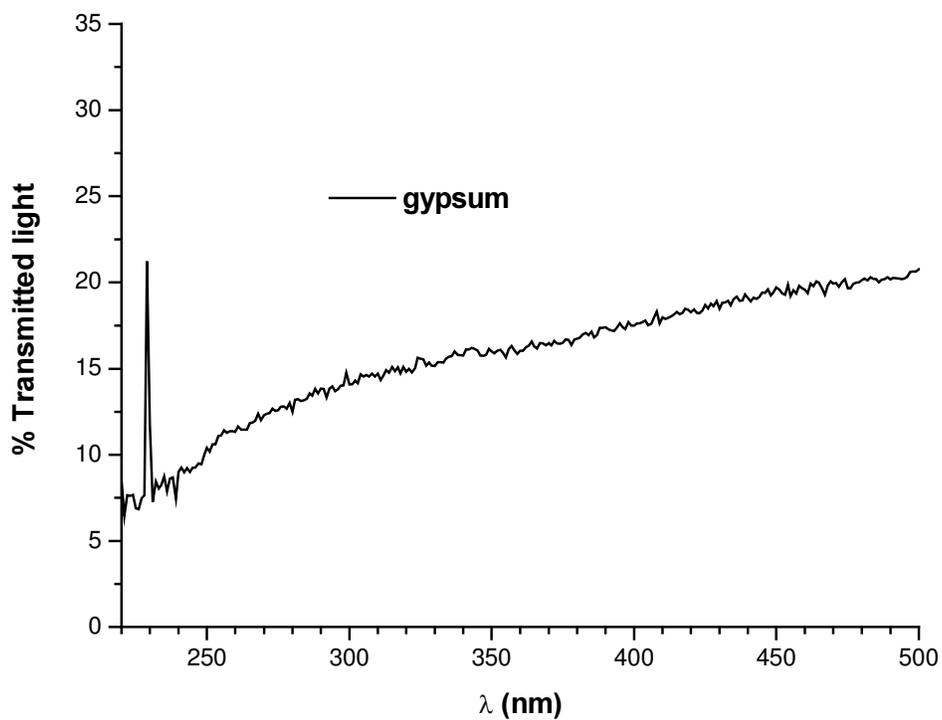

Figure 2: Light transmission spectrum (expressed as % transmitted light) of jarosite (A) vs. gypsum (B) samples.